\documentclass[10pt, letterpaper]{article}

\usepackage[a4paper, total={6.5in, 8in}]{geometry}
\usepackage{graphicx}
\usepackage{xcolor}
\usepackage{ragged2e}
\usepackage{placeins}
\usepackage[noadjust,compress]{cite}

\usepackage{graphicx}% Include figure files
\usepackage{dcolumn}% Align table columns on decimal point
\usepackage{bm}% bold math
\usepackage{float}

\newcommand{\beginsupplement}{%
        \setcounter{table}{0}
        \renewcommand{\thetable}{S\arabic{table}}%
        \setcounter{figure}{0}
        \renewcommand{\thefigure}{S\arabic{figure}}%
     }

\begin{document}

\title{Scale-specific dynamics of large-amplitude bursts in EEG capture behaviorally meaningful variability}
\maketitle

\begin{large}

\noindent Kanika Bansal\textsuperscript{1,2,*},
Javier O. Garcia\textsuperscript{1,3},
Nina Lauharatanahirun \textsuperscript{1,4},
Sarah F. Muldoon\textsuperscript{5},
Paul Sajda\textsuperscript{2,6},
Jean M. Vettel\textsuperscript{1,3,7}\\
\end{large}

\noindent\textsuperscript{1} U.S. Combat Capabilities Development Command Army Research Laboratory, Aberdeen Proving Ground, MD 21005, USA\\
\textsuperscript{2} Department of Biomedical Engineering, Columbia University, New York, NY 10027, USA\\
\textsuperscript{3} Department of Biomedical Engineering, University of Pennsylvania, Philadelphia, PA 19104, USA\\
\textsuperscript{4} Annenberg School of Communication, University of Pennsylvania, Philadelphia, PA 19104, USA\\
\textsuperscript{5} Mathematics Department, CDSE Program, and Neuroscience Program, University at Buffalo, SUNY, Buffalo, NY 14260, USA\\
\textsuperscript{6}Data Science Institute, Columbia University, New York, NY 10027, USA\\
\textsuperscript{7} Department of Psychological and Brain Sciences, University of California, Santa Barbara, CA 93106 , USA\\

\noindent \textsuperscript{*}Correspondence should be addressed to K.B. (phy.kanika@gmail.com).

\begin{abstract}
\noindent Cascading large-amplitude bursts in neural activity, termed avalanches, are thought to provide insight into the complex spatially distributed interactions in neural systems. In human neuroimaging, for example, avalanches occurring during resting-state show scale-invariant dynamics, supporting the hypothesis that the brain operates near a critical point that enables long range spatial communication. In fact, it has been suggested that such scale-invariant dynamics, characterized by a power-law distribution in these avalanches, are universal in neural systems and emerge through a common mechanism. While the analysis of avalanches and subsequent criticality is increasingly seen as a framework for using complex systems theory to understand brain function, it is unclear how the framework would account for the omnipresent cognitive variability, whether across individuals and/or tasks. To address this, we analyzed avalanches in the EEG activity of healthy humans during rest as well as two distinct task conditions that varied in cognitive demands and produced behavioral measures unique to each individual. In both rest and task conditions we observed that avalanche dynamics demonstrate scale-invariant characteristics, but differ in their specific features, demonstrating individual variability. Using a new metric we call \emph{normalized engagement}, which estimates the likelihood for a brain region to produce high-amplitude bursts, we also investigated regional features of avalanche dynamics. Normalized engagement showed not only the expected individual and task dependent variability, but also scale-specificity that correlated with individual behavior. Our results suggest that the study of avalanches in human brain activity provides a tool to assess cognitive variability. Our findings expand our understanding of avalanche features and are supportive of the emerging theoretical idea that the dynamics of an active human brain operate close to a critical-like region and not a singular critical-state.    

\end{abstract}

%\begin{keyword}
%Neural avalanches\sep Power-law \sep Electroencephalography
%\sep Emotional image viewing \sep Task-evoked functional reorganization 

%\end{keyword}

\section{Introduction}
Cognition is believed to require a widespread coordination of spatiotemporal neural activity. Though such coordination appears to be ubiquitous across tasks and conditions, the underlying principles of how this coordination occurs and how it may relate to individualized behavior are not yet well understood. Previous studies have proposed that cascading large-amplitude bursts in neural activity, also known as avalanches, provide a novel marker of characteristic complex dynamics that relate to brain function \cite{Beggs2003,Fukunaga2006,Tagliazucchi2012,Shriki2013,Arviv2015,Arviv2016,Priesemann2013,Shew2015,Palva2013}. In terms of analysis of avalanches in the human brain activity, the focus has been on \emph{resting} state activity, where several groups have demonstrated power-law probability distributions of avalanche sizes and durations \cite{Tagliazucchi2012,Shriki2013,Meisel2013,Priesemann2013}. Power-law distributions are interesting in that they imply the absence of a characteristic scale of activity -- i.e., scale-invariance. Systems with scale-invariant characteristics demonstrate efficient integration and segregation of information both locally and globally. Another observed attribute of neural avalanches has been that their branching process -- i.e., how the dynamics evolve over time -- demonstrates a balance between ongoing and upcoming neural activity \cite{Lombardi2012}, which is thought to be indicative of a macroscopic balance in excitation and inhibition \cite{Poil2012}. Interestingly, these statistical features of avalanche dynamics in resting state at the macro-scale (i.e., measured  fMRI, MEG, and EEG) were found consistent with what has been observed in smaller scale neuronal assemblies \cite{Beggs2003,Mazzoni2007,Pasquale2008,Friedman2012} and in vivo studies \cite{Petermann2009,Gireesh2008,Shew2015,Ponce-Alvarez2018,Miller2019}. These observations of brain dynamics, based on analyses adopted from statistical physics \cite{Sethna2001,Beggs2012}, have been the basis for the hypothesis that the brain operates near criticality \cite{Linkenkaer-Hansen2001,Beggs2003,Cocchi2017}, a special point in system's dynamical phase space, that separates order from disorder, providing for scale-invariance and subsequently, dynamical and functional diversity \cite{Haldeman2005,Beggs2008,Shew2009,Shew2013}. 

Despite these observations that critical-like features appear in experimental recordings across systems, questions surrounding the `criticality' hypothesis have not been fully explored. For example, the effect of stimulus and task-evoked activity on avalanche dynamics has only been partially investigated \cite{Arviv2015,Yu2017} and is not well understood \cite{Papo2014}. Additionally, the proposition that a single universality class exists and serves as a unifying mechanism for observed scale-invariance and criticality \cite{Friedman2012,Fontenele2019} theoretically require identical scaling features and a single critical point for all neural systems. However, given the variability of neural systems, compared to more traditional physical systems studied using this analytical approach (e.g. magnetic systems) \cite{Yeomans1992}, the `universality' proposition fails to provide an explanation for the  variability often observed in the study of neural avalanches \cite{Hahn2017,Dehghani2012,Yaghoubi2018}. Existence of criticality requires a delicate balance in system's dynamics and a single critical point requires a fine tuning of parameters to achieve this balance. Such a fine tuning appears unlikely in a complex system like the brain, considering the inherent nonstationarity of brain processes \cite{Kaplan2005} and the inhomogeneity of neuronal elements \cite{Moretti2013}. Moreover, a global organization of avalanches, probed by studying the probability distribution of avalanche sizes and durations, does not provide insight into the spatiotemporal cascading dynamics itself, which is of neuroscientific importance.

Here, we aim to address several of these issues by asking how previously observed scale-invariant, near-critical dynamics of avalanches in the `resting' brain (i.e., no controlled task) is related to state changes during stimulus-driven cognitive processing. We examined avalanches of neuronal activity from multi-channel scalp EEG, while participants underwent three sequential experimental conditions (Figure 1A). The conditions systematically varied in cognitive complexity, ranging from (i) resting state (eyes open), (ii) passive viewing of emotionally charged images, and (iii) active viewing of negatively charged images before rating the emotional intensity. Emotional responses are ubiquitous in our everyday lives and have been shown to affect brain activity by predominantly engaging neural activity in frontal and parietal regions of the brain at distinguishable timescales \cite{Buhle2014,KOHN2014}. Our relatively simple experimental design allowed us to investigate avalanche properties as a function of cognitive complexity, and enabled us to consider aspects of the `criticality' hypothesis as it relates to behaviorally meaningful variability.

\section{Methods}
\subsection{Participants}
36 healthy adults were recruited with average age $32.2\pm 7$. This study was carried out in accordance with the accredited Institutional Review Board at US Army Research Laboratory and conducted in compliance with the US Army Research Laboratory Human Research Protection Program (32 Code of Federal Regulations 219 and Department of Defense Instruction 3216.01). All participants gave informed, written consent. 

\subsection{Experimental design}
Participants performed three sequential experimental conditions (Figure 1A). The conditions systematically varied in cognitive complexity, ranging from (i) resting state (eyes open) with no explicit task, (ii) passive viewing of emotionally charged images with no explicit judgment, and (iii) active viewing of emotionally  charged images before rating the emotional intensity. For passive viewing, subjects viewed images with positive, negative, and neutral valence. For the active viewing task, only negatively charged images were shown. Participants used the numeric keypad on the keyboard to rate their emotional intensity in response to each image, for a total of 60 images, on a scale ranging from 1 (low emotional intensity) to 9 (high emotional intensity). Experimental timelines for these tasks are shown in Figure 1A and further details can be found in reference \cite{Roy2019}. 

\subsection{EEG data acquisition and pre-processing}
Continuous EEG recordings were captured via the Biosemi ActiveTwo EEG system (Amsterdam, Netherlands) equipped with standard Ag/AgCl electrodes from 64 sites on the scalp. Reference electrodes were placed on the mastoids. VEOG and HEOG external electrodes were used around the eyes during passive viewing and active viewing to ensure that our analysis was not affected by eye-blinks (see Supplementary Figure \ref{fig:removing_blinks}). Raw EEG measurements were pre-processed using in-house software in MATLAB (Mathworks, Inc., Natick, MA, USA) and the EEGLAB toolbox \cite{DELORME2004}. The pre-processing pipeline consisted of four steps (the PREP approach, \cite{Bigdely-Shamlo2015}): (1) resampling the raw EEG to 256 Hz; (2) line noise removal via a frequency-domain (multitaper) regression technique to remove 60 Hz and harmonics present in the signal; (3) a robust average reference with a Huber mean; and (4) artifact subspace reconstruction to remove residual artifact (the standard deviation cutoff parameter was set to 10). 

\subsection{Identification of avalanches in the EEG activity}
In microscopic brain imaging data, avalanches are identified as spatiotemporal clusters of events which are separated by windows of inactivity \cite{Arviv2015,Meisel2013,Priesemann2013}. Here, using 3 minutes of recorded data for each participant and condition, we first identified events as positive and negative excursions beyond a chosen threshold of 3 standard deviations for each EEG sensor (Figure 1B) \cite{Shew2015,Shriki2013}. Next, to identify avalanches, we looked for a continuous cascade of events across all sensors such that all the consecutive events are separated by a time span not larger than the correlation window $\Delta t$ (Figure 1C). The correlation window was selected in an adaptive manner for each individual by calculating the average inter-event-interval for all the consecutive events that occurred across all the sensors \cite{Shew2015,Shriki2013,Priesemann2013}. The observed mean $\Delta t$ was $28\pm4$ ms. The size of an avalanche ($S$) was determined as the total number of events within the avalanche. Avalanches of the same size can have different spatiotemporal spreads, as demonstrated by the orange cluster and the purple cluster in Figure 1C, (both have a size $S=11$).

\subsection{Power-law fitting and quality of fit}
To evaluate the global-scale (whole-brain) organization of avalanches, we used a maximum likelihood method to fit a head and tail truncated power-law to the avalanche distributions as described previously \cite{Klaus2011,Shew2015}. The fitting function used was $P(S)=S^{-\tau}(\sum_{x_{min}}^{x_{max}} x^{-\tau})^{-1}$. Here, $\tau$ is a fitting parameter and is the exponent of the power-law. For fitting $\tau$, values between 1 and 5 were tried with a step size of 0.01. Another fitting parameter is $x_{min}$ which represented the lower bound of the fit. We restricted its value to be $<15$. Finally, $x_{max}$ here denotes the upper bound of the fit and its value was chosen as the maximum avalanche size for which the avalanche size distribution showed a significant fit to a power-law.  
To ensure the validity of the fitted power-law, we tested the quality of the fit by computing the quality factor $q$ \cite{Clauset2009,Shew2015,Touboul2010}. For this computation, we used an established approach \cite{Clauset2009} and constructed $1000$ synthetic data sets which had the same number of observations as the fitted experimental data and were drawn from an ideal power-law with the same fitting parameters as the experimental data. We then calculated Kolmogorov-Smirnov (KS) statistics between synthetic data sets and their own power-law fit models. $q$ is the fraction of these synthetic KS-statistics which were greater than the KS-statistics for the experimental data. A power-law fit was deemed significant with a conservative criterion of $q\geq0.1$ \cite{Clauset2009,Shew2015}.

% figure 1
\begin{figure*}[ht]
\centering
\includegraphics [width=0.7\linewidth, keepaspectratio]
{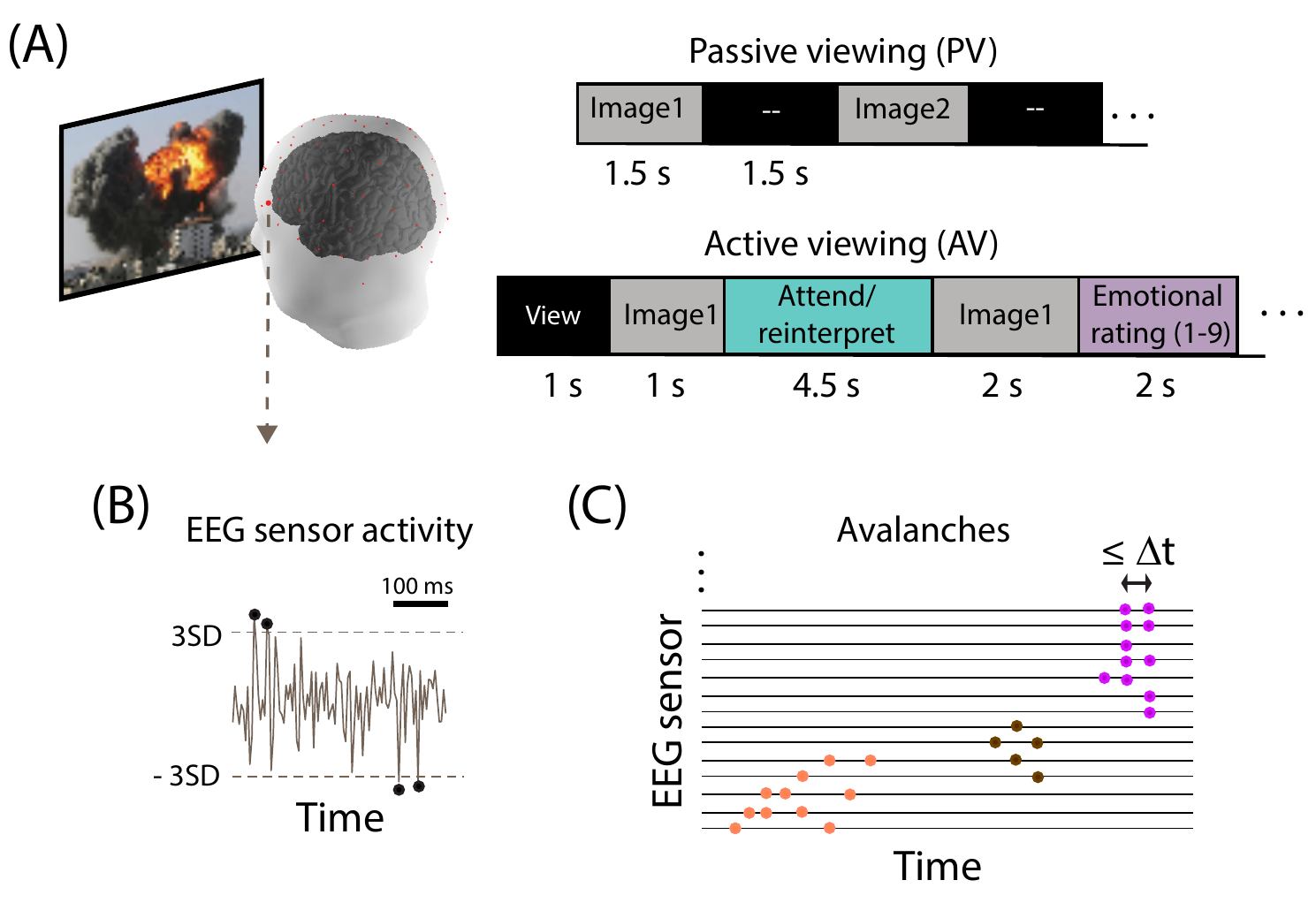}
\caption{Research design. (A) EEG was recorded during three different conditions that systematically varied in cognitive complexity. These conditions included a resting state and two task states which were passive viewing (PV) of emotionally charged images with no explicit judgment and active viewing (AV) of negatively charged images before rating the emotional intensity on a scale from 1 (low) to 9 (high). Time line of tasks was as shown here. (B)-(C) Avalanches were extracted from the EEG recordings of each task. An avalanche was identified as a cluster of events (activity beyond $\pm3$ SD of the mean, black dots) across all sensors, such that the temporal gap between any two consecutive events was not more than the size of the correlation window $\Delta t$. Size ($S$) of an avalanche was defined as the number of events in the avalanche cluster. As shown in C, observed avalanches varied in their spatiotemporal spread.}
\label{fig:schematic}
\end{figure*}

\subsection{Calculating branching parameter}
We extracted another feature of avalanches termed the `branching parameter' ($\sigma$) which denotes the average ratio of the number of events in the second and the first half of the observed avalanches \cite{Beggs2003,Arviv2015}. For each participant and condition, $\sigma=1/N\sum_{i=1}^{N} n_{T2}^i/n_{T1}^i$. Here, $i$ represents an avalanche, $N$ is the total number of avalanches in the given data segment, $n$ represents the number of events, and $T2$ and $T1$ denote the second and the first half of a given avalanche.  

\subsection{Comparing observed avalanche features between conditions}
We used a paired t-test to assess if computed global-level avalanche features demonstrate significant group differences due to the change in experimental condition. These features included fitted power-law exponents and branching parameters.  

\subsection{Calculating normalized engagement}
We analyzed localized features of avalanche dynamics by calculating the `normalized engagement' (NE) of each EEG sensor in producing avalanches. We defined engagement as the average number of `events' that a given sensor contributes to observed avalanches. If $N_s$ is the total number of avalanches under consideration and $n_i$ is the total number of events observed on a sensor $i$ during these avalanches, the engagement of the sensor $E_{i}=n_i/N_s$. We computed normalized engagement by normalizing each $E_{i}$ by the maximum value of $E_{i}$ across sensors, and consequently, obtained values bounded between 0 and 1. For each subject and condition, we calculated NEs during all the observed avalanches as well as during avalanches with specific ranges of sizes namely \emph{short} (1-10 events), \emph{moderate} (11-100 events), and \emph{persistent} (101-1000 events). 

\subsection{Identifying task-evoked changes in avalanche dynamics}
We used NE to identify regions of interest (ROIs) which showed significant task-evoked changes in avalanche dynamics. ROIs were those EEG sensors for which the distributions of NE values were significantly different between rest and active viewing, as per a paired t-test (uncorrected). To extract ROIs for the regression model (see below) and figures presented here, we used a significance level ($\alpha$) of 0.05. However, to test the robustness of our findings, different $\alpha$ values (between 0.02 and 0.07) were used.

\subsection{Statistical analysis}
In order to test a relationship between avalanche dynamics and emotional ratings reported during the active viewing task, multiple regression analyses were conducted. The emotional rating ($Y$) for each participant was calculated as the average rating reported across all trials. For predictors in the regression models, we used average NE values across the identified ROIs for different ranges of avalanche sizes i.e., for short avalanches ($R_s$), for moderate avalanches ($R_m$), and for persistent avalanches ($R_p$). Prior to multiple regression analysis, all the predictors were mean-centered and tested to be normally distributed using the Shapiro-Wilk's test ($p>0.02$). In order to test the independent and interactive effects among the predictors on emotional ratings, main effects, pairwise interactions, and the three-way interaction term were included in the model as shown below.

\begin{equation}
Y = 1 + R_p + R_m + R_s + R_p*R_m + R_p*R_s + R_m*R_s \\
+ R_p*R_m*R_s. 
\label{eqnGLM}
\end{equation}

To probe the interaction effects for observed significant interactions, we conducted simple slope analyses to compare high ($>0.5$ SD) and low ($<-0.5$ SD) values of the moderator in relation to the mean.

\section{Results}

\subsection{Global-scale avalanche features vary between individuals and tasks} 
In our first analysis, we examined how the global, whole-brain functional activity, is potentially represented through EEG avalanche dynamics, and varies across our three experimental conditions. Previous findings have shown that the probability distribution of avalanche sizes during resting state fits a power-law, such that $P(S)\sim S^{-\tau}$, where $\tau$ is a positive valued exponent \cite{Shriki2013,Priesemann2013,Tagliazucchi2012,Meisel2013}. We examined if the distributions of avalanche sizes, derived from EEG, fit the power-law behavior both for resting state as well as the two task conditions. As shown in Figure 2A, in our data (circles), avalanche sizes spanned a little over two orders of magnitude; therefore, we fit the data to a truncated power-law (see Methods) \cite{Clauset2009,Klaus2011,Yu2014,Shew2015}. A fitted power-law is shown by the dashed line. Here, $x_{min}$ and $x_{max}$ represent the lower and upper bounds of the fit, describing the minimum and maximum values of avalanche size that can be fitted through power-law probability dynamics. To test the goodness of fit \cite{Touboul2010}, for each participant and each condition, we calculated the quality of the fit factor ($q$) as described by Clauset et al. \cite{Clauset2009} (see Methods). We used a conservative criterion and deemed the fit significant if $q\geq0.1$ \cite{Clauset2009,Shew2015}. 

% Figure 2
\begin{figure*}[t]
\centering
\includegraphics [width=1\linewidth, keepaspectratio]
{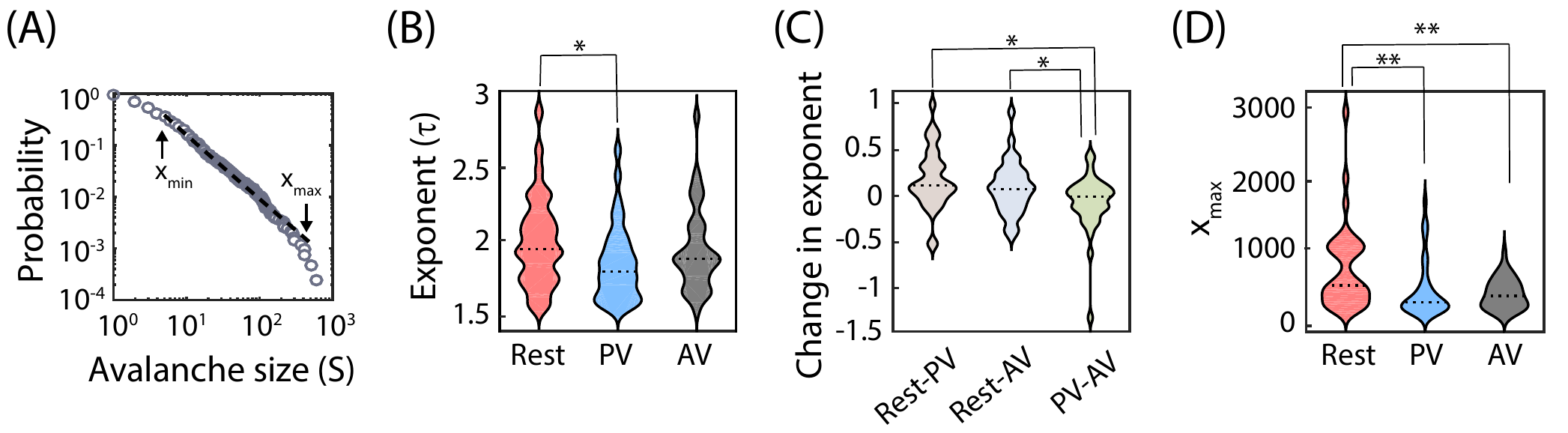}
\caption{Probability distributions of avalanche sizes fit to power laws. (A) We tested if the probability distribution of avalanche sizes for a given subject and condition fits a power-law such that $P(S)\sim S^{-\tau}$. In the example here, the circles show the cumulative probability distribution of avalanche size for the resting state condition of one participant. The dotted line shows the fitted power-law. Here, $x_{min}$ and $x_{max}$ denote the lower and upper bound of the fit. (B) We observed distributed values of fitted power-law exponent ($\tau$) for different participants across all conditions. We also observed a decreasing trend in exponent values with increasing cognitive complexity of task conditions. (C) Pairwise difference in exponent values between conditions showed distributed non-zero values, indicating within subject variability in exponents, and therefore avalanche distributions, across task conditions. Difference of exponents is significantly more prominent between resting and task conditions as compared to the difference between PV and AV conditions. (D) We observed a significant decrease in $x_{max}$ for the two task conditions as compared to the rest, indicating that large size avalanches during task conditions significantly deviate from their resting state distributions. Here, * and ** denote a significant difference on paired t-test with \textit{p}-value $\leq$0.01 and $\leq$0.001 respectively.}
\label{fig:variability}
\end{figure*}

We observed a distribution of the fitted power-law exponent $\tau$ across participants for all three conditions, as shown in Figure 2B (individual values for $\tau$ and $q$ are shown in Supplementary Figure \ref{fig:individual_exponents}). While across participants and conditions we found the avalanche distributions to fit a power-law, the exponent significantly varied between individuals. As shown in Figure 2B, exponents for the resting task varied between the range $\sim1.5$ to $\sim3$ across subjects. A similar variability was observed for the passive viewing and active viewing conditions. We also observed a slight decreasing trend in the exponent values for increasing task complexity, with a significant difference between rest and passive viewing ($t(34)=2.84$, $p=0.008$). 

In Figure 2C, we plot the pairwise difference in exponents within an individual for each pairing of the experimental conditions. Notably, all of the conditions show non-zero differences. This variability in the exponent suggests that the global scale-invariant organization may shift based on the cognitive complexity of the task. An interesting trend was a decrease in the change in exponents as the task complexity increased -- i.e., we observed significantly higher differences between rest and task conditions (Rest-PV, Rest-AV) than between the two task conditions (PV-AV) ($t(34)=2.62$, $p=0.01$; $t(34)=2.84$, $p=0.008$).  

In addition to the variability observed in the exponent values of the power-law fits, we also observed that $x_{max}$, which represents the upper bound of the fit, decreases from rest to task conditions, seen in  Figure 2D (rest to PV $t(34)=3.63$, $p<0.001$ and rest to AV $t(34)=3.75$, $p<0.001$; also see Supplementary Figure \ref{fig:combined_cdfs}). Collectively, these results indicate that EEG-derived avalanches that are spatially and temporally distributed, demonstrate significant power-law dynamics for both rest and our two task conditions, though the form of the power laws varies significantly across both individuals and conditions.

\subsection{Avalanche dynamics vary between individuals and task conditions, but in a correlated manner}
We computed the branching parameter ($\sigma$) to further assess EEG-derived avalanche dynamics and investigate whether the system was in a state of criticality. Branching parameter values demonstrated similar variability as the power-law exponents, both across participants and conditions. As shown in Figure 3A, we observed that the branching parameter show significant increase from the rest to passive viewing ($t(35)=-2.76$, $p=0.009$) and to active viewing ($t(35)=-2.99$, $p=0.005$). The medians of these values in each experimental condition are close to the theoretical prediction of 1, suggesting the system is close to a critical state \cite{Priesemann2014}. 

% Figure 3
\begin{figure*}[t]
\centering
\includegraphics [width=1\linewidth, keepaspectratio]
{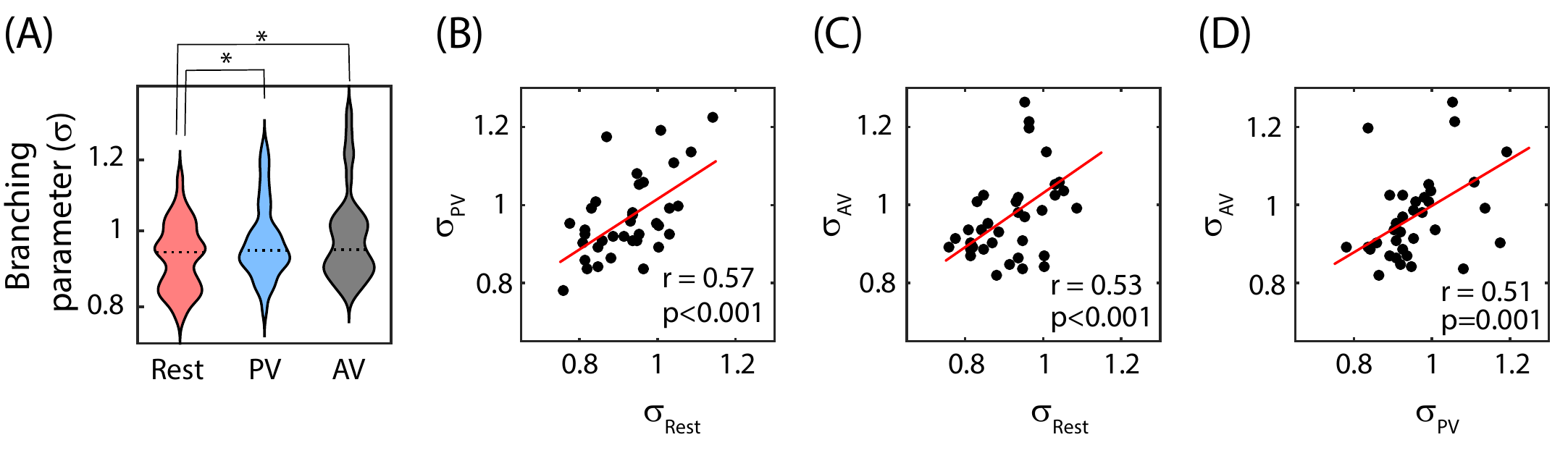}
\caption{Estimated branching parameters. (A) The branching parameter ($\sigma$) varies across participants and conditions and indicates an increasing trend with increasing cognitive complexity. Here, * denotes a significant difference on paired t-test with \textit{p}-value $\leq$0.01. (B)-(D) Despite the variability within and between participants, we observed a significant positive correlation in branching parameter between conditions. Here, $r$ denotes the Pearson's correlation coefficient and $p$ denotes the associated \textit{p-value}.}
\label{fig:sigma_correlations}
\end{figure*}

While we observe variability in the values of the branching parameter across individuals, there are clearly significant correlations within an individual between parameters across different experimental conditions. Figures 3B-D compare within-individual branching parameter values between conditions. For all pairings of conditions, we found a significant positive correlation between the branching parameter values (rest and passive viewing $r=0.57$, $p<0.001$; rest and active viewing $r=0.53$, $p<0.001$; passive viewing and active viewing $r=0.51$, $p=0.001$).  

Results thus far indicate a consistent interpretation, namely EEG activity shows evidence of the brain being at a near-critical state, as quantified via the observed power-law distributions and branching parameter values for the avalanche activity. However, neither of these measures suggests a universality class, and both reveal the sensitivity of these measures to individual differences between participants. 

% Figure 4
\begin{figure}[ht]
\centering
\includegraphics [width= 0.55\linewidth, keepaspectratio]
{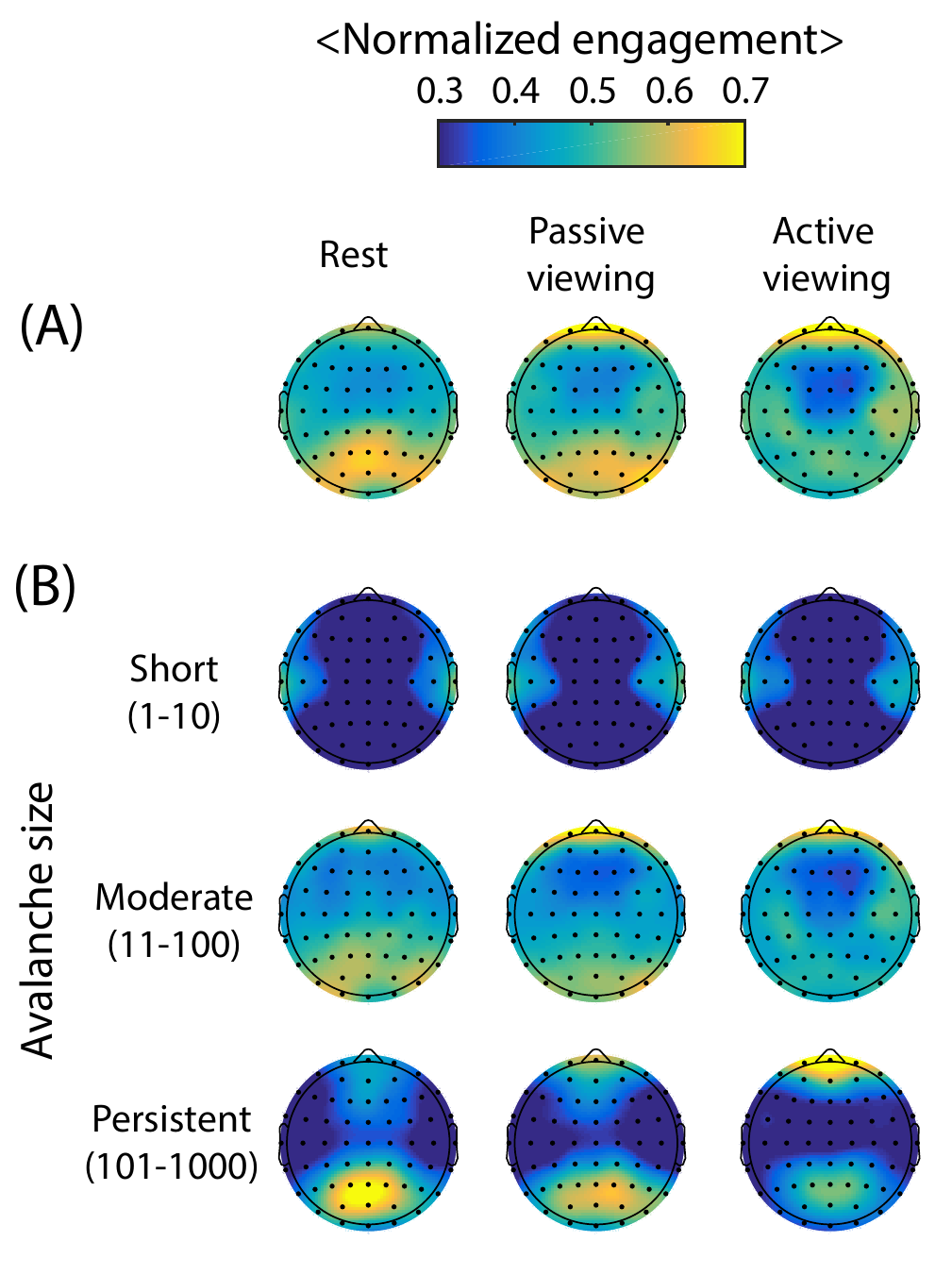}
\caption{Localized avalanche features change with task conditions. We calculated normalized engagement (NE) as the normalized average frequency of each sensor for contributing events to observed avalanches. (A) Subject averages of NE (denoted by $<>$) for each sensor across different conditions show a systematic shift from parietal to frontal activity as cognitive complexity increases (i.e. from resting state to active viewing). (B) NE shifts are specific to avalanche size. Subject averages for short avalanches  (1-10 events) show similar topologies for all three conditions. However, one sees a shift in the topology between conditions for moderate (11-100 events) and persistent (101-1000 events) avalanches.}
\label{fig:patterns}
\end{figure}

\subsection{Changes in  task complexity result in localized changes in avalanche dynamics}

We next examined whether avalanche dynamics, defined locally, relate to the task complexity that varies across experimental conditions. Specifically, we characterized the EEG sensors located at spatially distinct positions on the scalp using a  metric we term `normalized engagement' (NE, see Methods). Larger NE values indicate more frequent high amplitude activity on the sensor. 

We observed differences in the spatial distribution of large NE values across  the three experimental conditions. Figure 4A, shows the group average of these distributions for three task conditions. We find that larger NE values systematically shifted from posterior areas of the scalp to frontal areas as the task complexity increased from resting state to passive and then active viewing.

Next we considered these spatial distributions as a function of avalanche size. We binned avalanches into three scales based on their sizes (short, moderate and persistent) and recomputed the NE measures as a function of task complexity. Figure 4B displays these results and shows that different avalanche sizes are associated with a different spatial distribution of normalized engagement across the scalp. For short avalanche sizes, there is no substantial difference between the spatial distributions across task conditions. However, for persistent avalanches, there is a clear shift in the spatial distribution of large NE values when comparing the rest and task conditions. Importantly, the shift shows that as the task becomes more complex (active viewing) there is a substantial engagement in both frontal and parietal sensors. Additional analyses, investigating the frequency-specificity of these results, showed that the effects were sustained in EEG activity below 20 Hz, indicating they were unlikely linked to muscle artifact and EMG contamination \cite{Muthukumaraswamy2013,Goncharova2003}. More interestingly, for persistent avalanches, the spatial distribution disassociated across frequencies, with occipital patterns expressed in the alpha band activity (8-12 Hz) while the frontal patterns were largely in the theta band activity (4-7Hz) (see supplementary Figures \ref{fig:frequency_dependence}).  

Resting state EEG is characterized by strong alpha oscillations in occipito-parietal cortex \cite{Goldman2002,Basar2012}. Additionally, it  has been shown that visual stimuli and emotional judgments typically result in high theta activity in  frontal brain regions \cite{Cavanagh2014,Ertl2013}. Importantly, persistent avalanche dynamics provide links between frontal and occipito-parietal activity in a way that is consistent with other findings showing that the coordination of these networks is important for executive function and behavior \cite{KOHN2014,Buhle2014}.

\begin{table*}
\centering
\includegraphics [width=0.8\linewidth, keepaspectratio]
{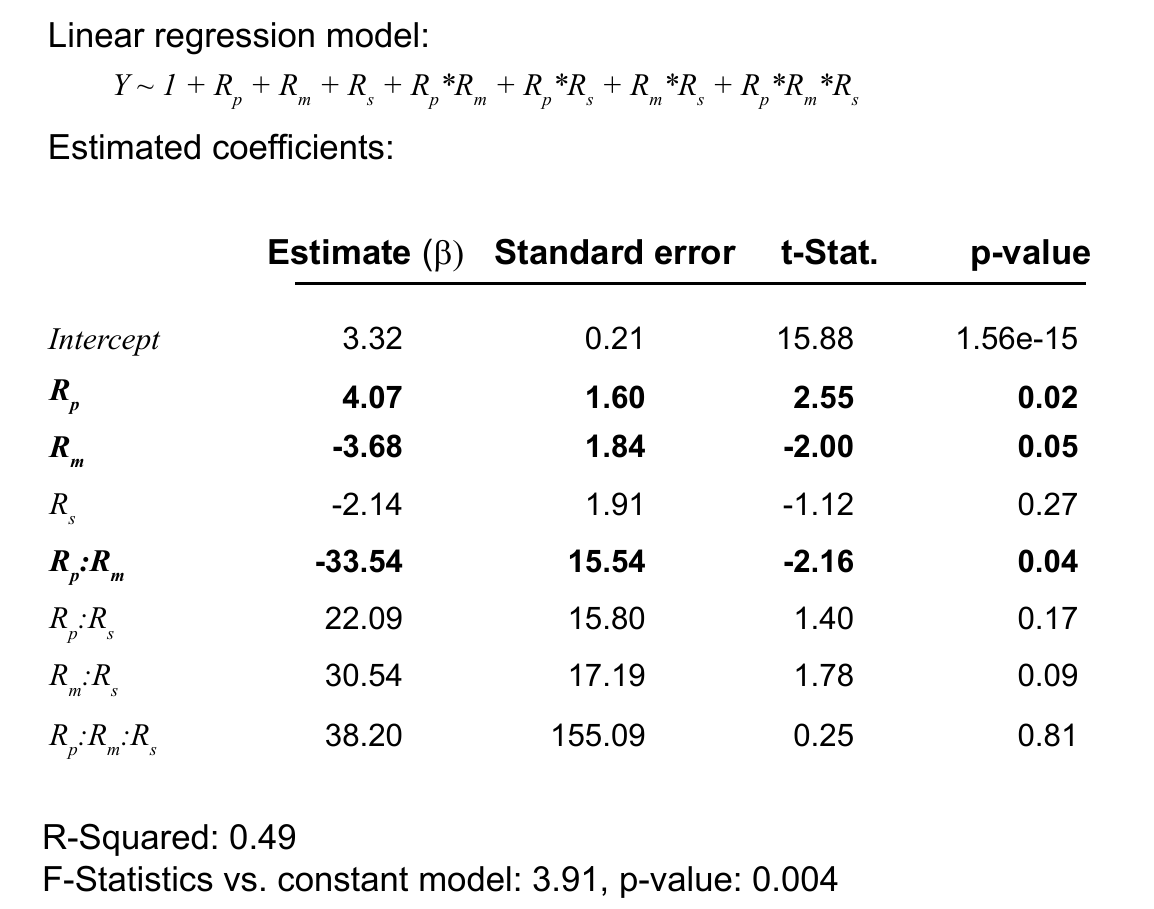}
\caption{Results from the regression model described by Equation \ref{eqnGLM} to assess the relationship between localized avalanche features and emotional ratings reported by the participants during active viewing task. Here, $Y$ represents the mean emotional rating for each individual, $R_p$, $R_m$, and $R_s$ represent average normalized engagement (NE) values of regions of interest for persistent, moderate, and short avalanches respectively. In the comprehensive model, we included main effects, pairwise interactions, and three-way interaction terms. Results for each model term are presented in different rows of the table and significant effects are highlighted in the bold text. An interpretation of model findings is discussed in the text and Figure 5.}
\label{fig:table}
\end{table*}

\subsection{Variability of task evoked localized responses in avalanche activity is predictive of individual behavioral response}

We used multiple regression analyses to investigate whether the localized variability in avalanche dynamics correlates with subject level differences in behavior, measured through self-reported emotional ratings during active viewing condition. EEG sensors which showed significant task-evoked change in NE values (using a paired t-test with $\alpha = 0.05$, see Methods) are highlighted in Figure 5A. We found 27 sensors for persistent avalanches, 23 for moderately sized avalanches (see Figure 5A), and only five sensors (two frontal and three parietal) for short avalanches (not shown in Figure 5A). We call these sensors `regions of interest' (ROIs).

% Figure 5
\begin{figure*}[ht]
\centering
\includegraphics [width=1\linewidth, keepaspectratio]
{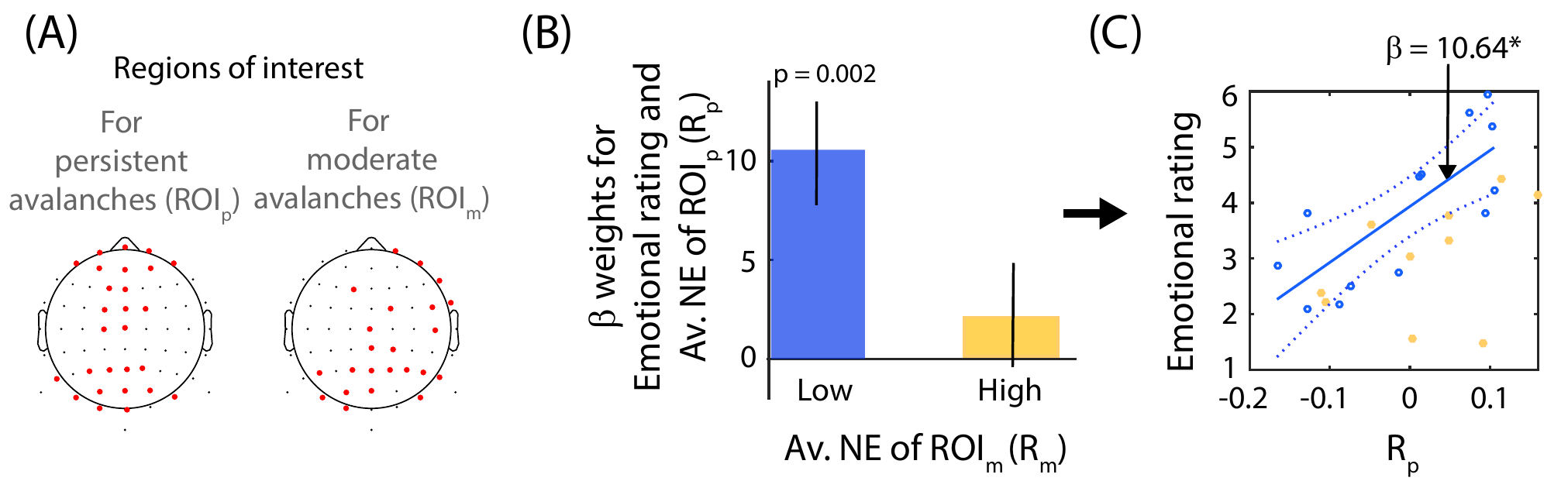}
\caption{Avalanche dynamics correlate with subject specific emotional ratings in active viewing task. (A) Set of regions (EEG sensors marked in red) which showed significantly different NE values between rest and active viewing for persistent (left) and moderate (right) avalanches. Using a regression model, we found that average NEs across these regions of interest ($R_p$ for persistent avalanches and $R_m$ for moderate avalanches), were predictive of self reported emotional ratings by participants. In our model, in addition to the main effects of $R_p$ and $R_m$ on emotional rating, we also found a significant interaction effect between these quantities. Details of the model are presented in Equation \ref{eqnGLM} and Table 1. In (B)-(C) we probe this interaction effect by assessing the relationship between emotional rating and $R_p$ (predictor) through linear regression for low and high values of the moderator ($R_m$). We observed a positive relationship between $R_p$ and emotional rating for low values of $R_m$ ($\beta = 10.64$, $p = 0.002$, $R^2=0.59$), whereas no significant relationship was found between $R_p$ and emotional rating for high values of $R_m$ ($\beta = 1.74$, $p = 0.47$). Here, error bars denote $\pm1$ standard error and dotted lines denote 95\% confidence interval.}
\label{fig:behavior}
\end{figure*}

We used a linear regression model (Equation \ref{eqnGLM}) to correlate emotional ratings ($Y$) with average NEs across ROIs for each scale of the avalanche dynamics ($R_p$ for persistent, $R_m$ for moderate, and $R_s$ for short), and the model findings are detailed in Table 1. We found three significant relationships: (i) a main effect showing that the average NE of the ROIs for persistent avalanche activity ($R_p$) is positively related to emotional rating ($\beta=4.07$ and $p = 0.02$); (ii) another main effect showing that the average NE of the ROIs for moderate avalanche activity ($R_m$) is negatively related to emotional rating ($\beta=-3.68$ and $p = 0.05$); and (iii) an interaction effect between ROIs engagement for persistent and moderate avalanche activity ($\beta=-33.54$ and $p=0.04$).

As shown in Figures 5B-C, to probe the observed significant interaction effect, we conducted an analysis of these slopes ($\beta$s), comparing high and low engagement (assessed by NE) of the ROIs in the avalanche activity in relation to mean. We observed that engagement of the ROIs in persistent avalanche activity shows a positive relationship with emotional rating when the engagement in moderate avalanche activity is low ($\beta=10.64$ and $p=0.002$, $R^2 = 0.59$). Whereas, this relationship disappears when the engagement in moderate avalanche activity is high ($\beta = 1.74$, $p = 0.47$). This suggests that localized avalanche dynamics capture spatiotemporial coordinated activity that are predictive of subject experience and behavior.

\section{Discussion}
We investigated avalanche features in the EEG time series as a way to characterize brain states across individuals and tasks of varying complexity. For the global organization of avalanches, we observed power-law behavior and scale-invariance in the resting state condition as well as in  the passive and active viewing conditions. These findings support previous studies that have investigated resting state brain activity \cite{Shriki2013,Priesemann2013,Tagliazucchi2012}, as well as  studies where there is a presence of a stimulus, both for humans \cite{Arviv2015,Palva2013} and non-human primates \cite{Tomen2014,Yu2017}. The results also suggest that the large-scale neural dynamics within the brain maintain a critical-like state \cite{Cocchi2017,Ponce-Alvarez2018,Linkenkaer-Hansen2001} for different conditions and that this state may be important for optimizing information processing capabilities within the brain \cite{Shew2009,Beggs2008,Haldeman2005,Papo2014}.  

Though the `critical-like' state was prevalent across conditions and subjects, the corresponding power-law exponents, ($\tau$), displayed substantial variability across individuals and between conditions within an individual. The estimated values in macroscopic experimental data deviate from the exponent value of $\tau=1.5$ reported in multiple organisms as well as  in in-vitro experiments \cite{Beggs2003,Hahn2010,Priesemann2013,Yu2017,Shew2015}. The estimated value of $\tau$ can vary as a function of study parameters, e.g., the distance between recording sensors, the chosen threshold for identifying large-amplitude events, and the size of the correlation window $\Delta t$ \cite{Petermann2009,Priesemann2013,Miller2019}. Since the first two parameters were fixed during the study, they did not contribute to the observed variability in exponent. The third parameter was adaptively defined for each individual and condition as the mean inter-event-interval \cite{Shew2015,Priesemann2013} which allowed a uniform interpretation of estimated avalanche features across individuals and conditions. Observed variability in the exponent can therefore be attributed to individual differences in avalanche dynamics.  

The absence of a single-valued exponent to describe the avalanche dynamics likely indicates that the scale-invariant organization we see in the EEG is not fine tuned to a single ``universal" state, but instead, varies based on the underlying functional dynamics \cite{Yaghoubi2018}. Therefore, our results suggest that models for criticality in neural systems may need to  account for robust individual and task related differences \cite{Moretti2013,Williams-Garcia2014,Priesemann2018}. For example, Moretti et al.  suggested that the inherent hierarchical-modular architecture and heterogeneity of cortical networks can replace a singular critical point by an extended critical-like region such that the exponents may vary to capture unique functional connectivity of different neural systems and cognitive states \cite{Dehghani2012,Yaghoubi2018}. Despite the variability we see, the observed exponent values do distribute within a similar range across the three experimental conditions, which further suggests a flexible, yet functionally ordered organization within the brain, characteristic of a complex system.

Surprisingly, concurrent with the maintained global scale-invariance, we observed localized features of avalanches to show not only a task-dependent regional patterning, but also the scale-specific features. We found that the avalanches with relatively large sizes (moderate and persistent) carried meaningful information about the task-dependent changes within the brain. Using a metric that quantifies the likelihood of different brain regions to engage in avalanches, we identified a set of ROIs which were predictive of individual behavior, i.e. emotional rating, through a regression model that also included interaction terms relating the different scales of avalanche sizes. We found these ROIs to be located within the frontal as well as occipito-parietal regions of the brain. Previous research has also highlighted the activation of these areas during the processing of emotional stimuli \cite{Baumgartner2005,KOHN2014,Buhle2014}. 

We used a data-driven approach to identify these ROIs and required that they show a significant change of their engagement in producing avalanches, following the change in experimental condition from rest to active viewing. To assess a significant change in engagement, we used a paired t-test with $\alpha = 0.05$. Importantly, the findings of our regression model were qualitatively robust for a range of $\alpha$ (0.02 to 0.07). We observed higher average engagement of ROIs in producing persistent avalanches to be indicative of higher emotional rating. 

Combining our findings across the global and regional scales of the EEG dynamics, we propose that the macro-scale dynamics of the neural activity operates close to criticality whilst simultaneously rapidly changing to match behavioral goals \cite{Karimipanah2017}. Neurophysiological mechanisms for the brain to adaptively maintain its macroscopic organization near criticality are not well understood. However, it appears likely that such an adaptive mechanism is characteristic of  healthy human brains \cite{Massobrio2015}. Indeed, it has been previously shown that the features of criticality are more consistently disrupted in the interictal activity of epilepsy patients \cite{Arviv2016}, where, a localized or global inability of the brain to regain its balance may cause uncontrolled activity or seizures. A significant departure from a critical-like state has also been reported during sustained wakefulness over several hours, which reverses back upon the recovery of sleep \cite{Meisel2013}, a necessity for the healthy brain function \cite{Banks2007}. Scale-invariant avalanche dynamics have also been found important during developmental stages of the brain \cite{Iyer2015} and during recovery after brain insult \cite{Roberts2014}.  

We believe that the analysis of macroscopic avalanches can provide useful insight into the temporal evolution of brain activity and might even provide biomarkers when the activity becomes abnormal. However, the study of avalanches within the human brain also presents a number of challenges. Avalanches estimated in EEG recordings allow only a limited neurophysiological interpretation since the EEG represents large population neural activity measured at the scalp, and suffers from volume conduction distortion and artifacts due to the high electrical conductivity of the scalp. Often, source localization is used to identify unique sources in the brain given the measured scalp EEG activity and to rid the data of the artifacts due to volume conduction. However, effective and accurate localization of signal sources is a field of research in itself and the analysis of avalanches in such data would require careful consideration and further work.

\subsection{Conclusions}
We observed that similar to the resting state activity, EEG-derived avalanches in task-evoked, stimulus-driven activity demonstrate power-law probability distributions and scale-invariance, characteristic to global organization in a complex system near criticality. Global avalanche dynamics however varied between individuals, as represented by the exponents of the fitted power-laws. From the analysis of the branching parameters, this global dynamics within an individual seemed to be correlated across experimental conditions. Analysis of localized (regional) avalanche dynamics showed correlation with behavioral measure, and  this correlation was driven by a spatiotemoporal interaction of the avalanches. In general, we believe the analysis of macroscopic avalanches in the EEG provides a straightforward yet effective way to measure the state of the brain, both in health and disease.

\section{Data availability}
Data is accessible upon request as far as allowed by the security policy and guidelines established with the ethics committee of the US Army Research Laboratory Human Research Protection Program. Requests should be addressed to K. Bansal.

\section{Author contributions}
K.B. conceived the idea of this research, implemented the analysis, prepared figures, and wrote the initial draft. J.M.V. secured the funding and contributed in performing the EEG experiments. J.O.G. helped with the EEG data pre-processing and data visualization. N.L. helped with the statistical modeling. P.S. and S.F.M. provided supervision. All authors contributed in reviewing and editing the manuscript. 

\section{Acknowledgment}
This research was sponsored by the US Army Research Laboratory and was accomplished under Cooperative Agreement Number W911NF-16-2-0158. The views and conclusions contained in this document are those of the authors and should not be interpreted as representing the official policies, either expressed or implied, of the Army Research Laboratory or the U.S. Government. The U.S. Government is authorized to reproduce and distribute reprints for Government purposes notwithstanding any copyright notation herein.

%\bibliography{references}

\newpage
\beginsupplement

\centering
\section*{{Supplementary information}}
\begin{figure*} [h]
\centering
\includegraphics [width=0.45\linewidth, keepaspectratio]
{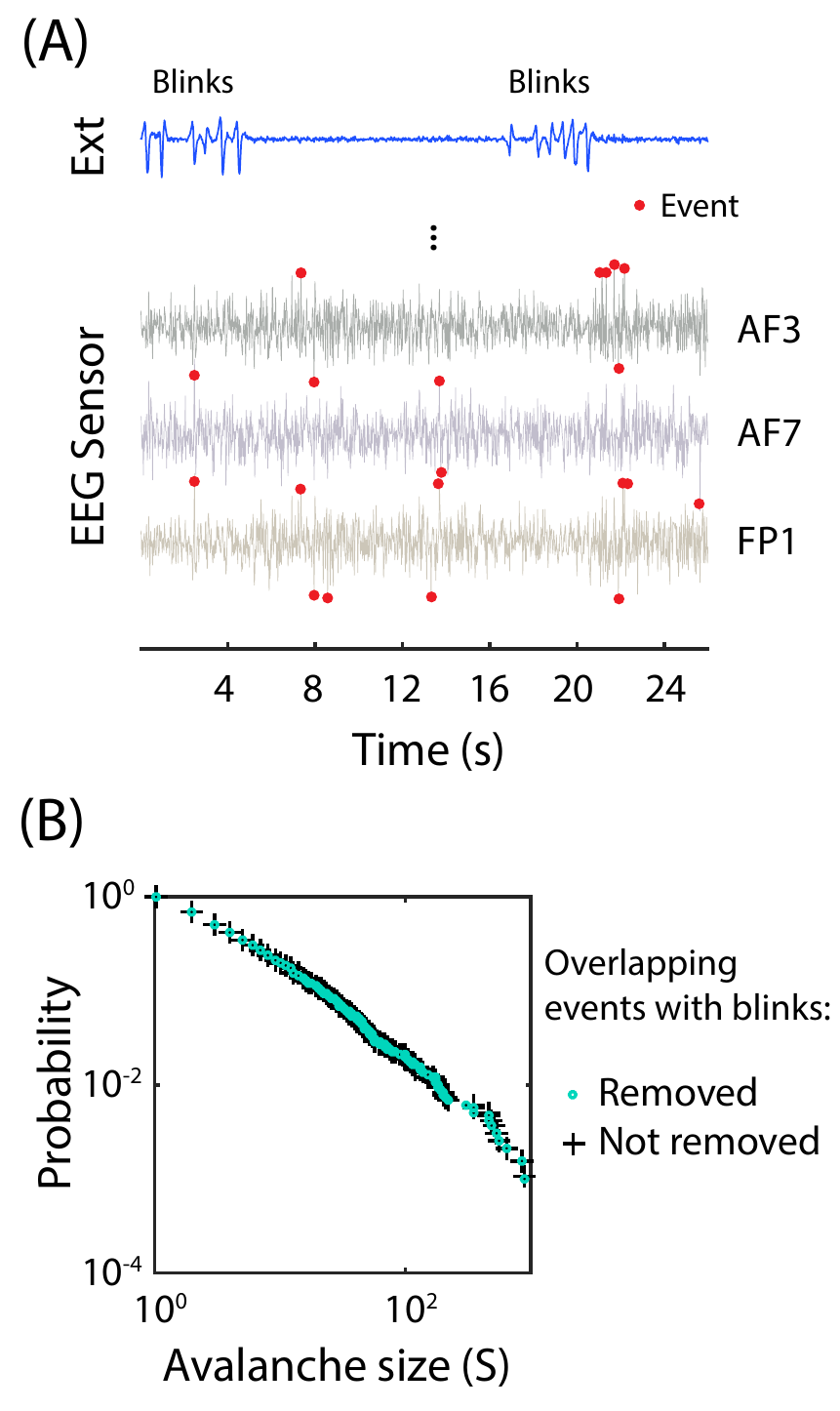}
\caption{During passive viewing and active viewing conditions, we monitored eye-activity from an external channel, placed close to the eyelid, which allowed us to specifically track eye-blinks and their impact on events and avalanches. (A) In addition to the EEG data preprocessing steps, in order to ensure there are no eye-blinks related artifacts in event distributions, we removed all the events in the frontal channels that showed an overlap with an eye-blink. (B) In general, global probability distribution of avalanche sizes remained largely unchanged before and after the removal of events which overlapped with eye-blinks.}
\label{fig:removing_blinks}
\end{figure*}

\begin{figure*} 
\centering
\includegraphics [width=1\linewidth, keepaspectratio]
{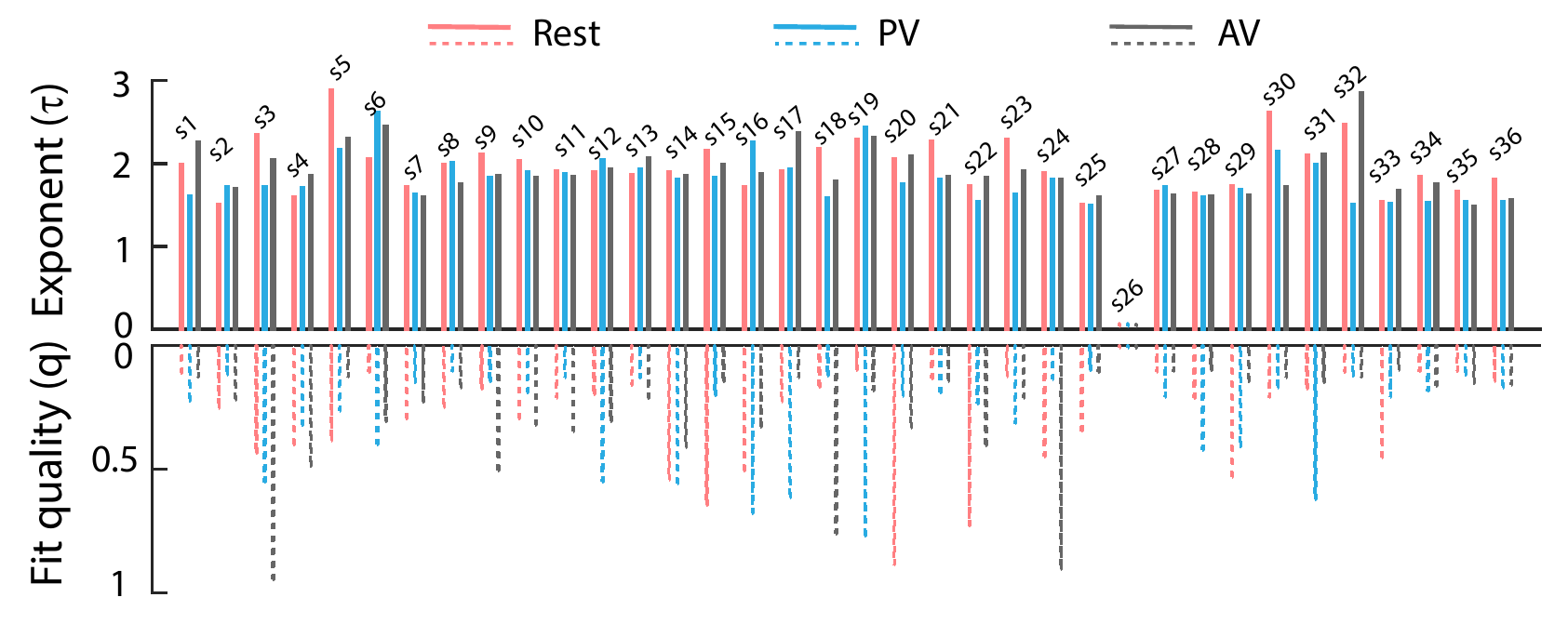}
\caption{We tested if avalanche size distributions fit to the power-law behavior for each participant and condition. To assess the goodness of a power-law fit, we calculated quality of the fit factor ($q$). We deemed the fit significant if $q\geq0.1$. Here, we show the fitted power-law exponent ($\tau$) and fit quality factor obtained for different participants and conditions. A significant power-law fit was obtained for all the participants expect one (s26), whose data did not show a significant fit within the tested range of parameters under our conservative significance criterion.}
\label{fig:individual_exponents}
\end{figure*}

\begin{figure*}
\centering
\includegraphics [width=0.6\linewidth, keepaspectratio]
{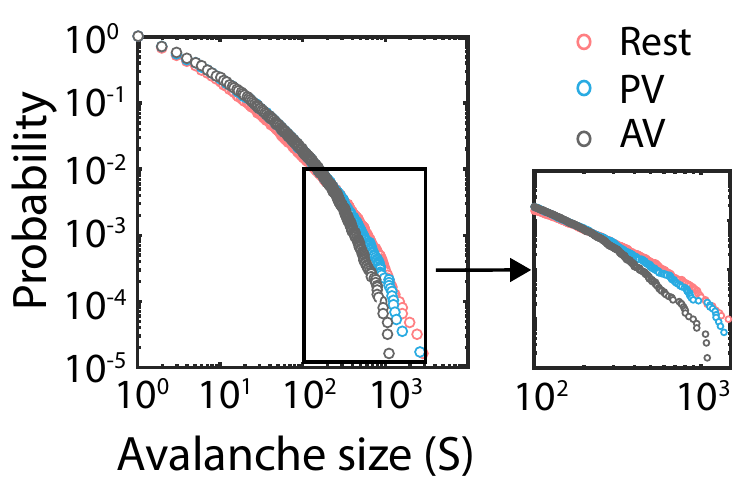}
\caption{Probability distributions of avalanche sizes calculated for avalanches observed across all the participants for three different conditions. Distributions show more pronounced differences between conditions for avalanches of relatively higher sizes.}
\label{fig:combined_cdfs}
\end{figure*}

\begin{figure*}
\centering
\includegraphics [width=1\linewidth, keepaspectratio]
{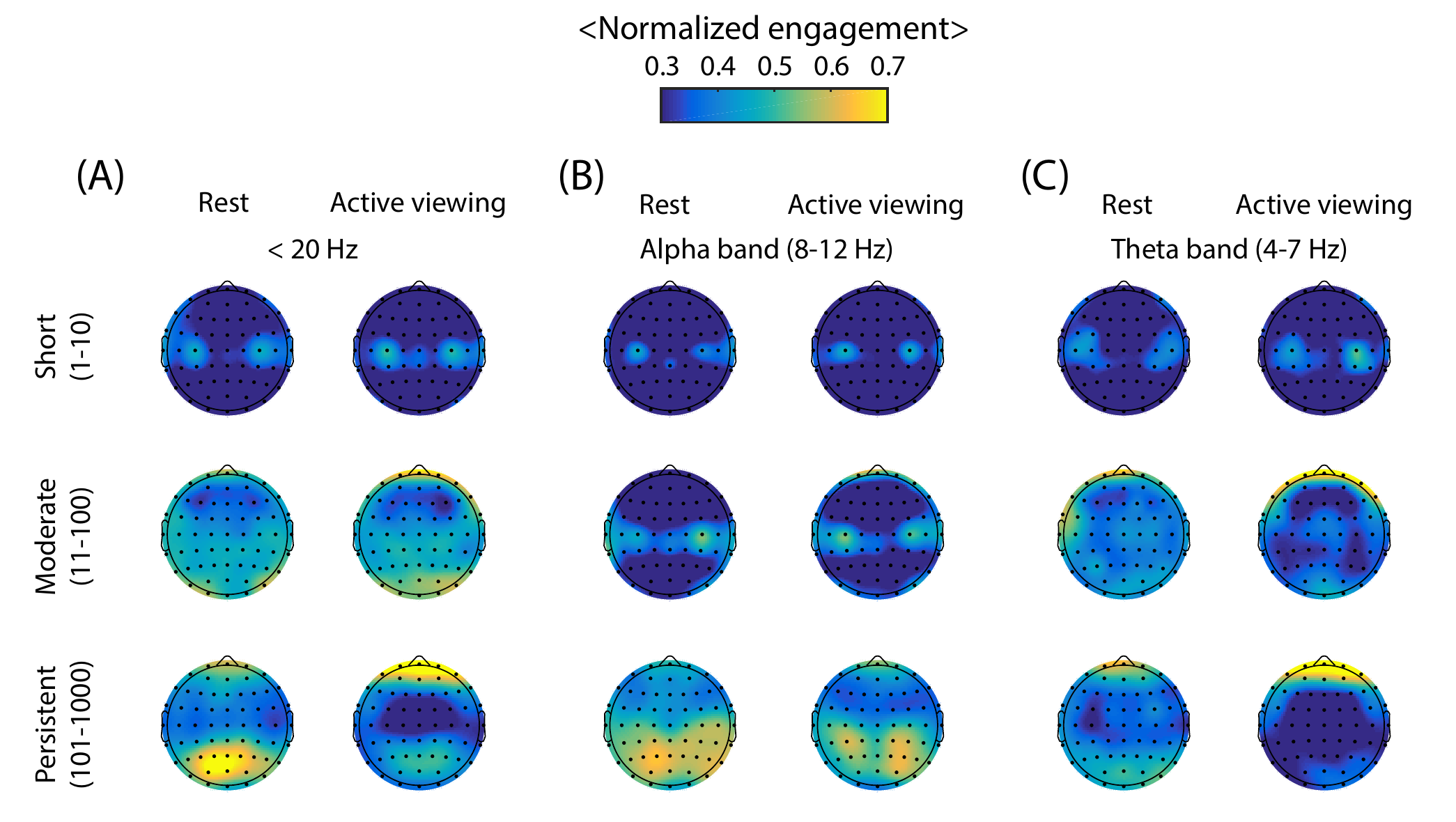}
\caption{Scale dependence of normalized engagement (NE) of EEG sensors in avalanches for filtered EEG data. In order to understand the change of NE values with changing task conditions, we compared these results within a more traditional EEG analysis framework, and analyzed NEs for EEG activity extracted in particular frequency ranges that are shown to have distinct functional purposes. (A) First, in order to ensure that the elevation in frontal NEs during task performance condition is not due to any muscle or EMG artifacts and represents neuronal responses, we filtered out the high frequency activity from EEG data that carries the largest impact of such artifacts \cite{Muthukumaraswamy2013,Goncharova2003}. We applied a 20 Hz lowpass filter ($6^{th}$ order Butterworth) to the EEG signal and then calculated NE values for the filtered data. We still observed similar frontal and occipito-parietal changes of NE values between rest and active viewing task as discussed in the context of Figure 4. (B)-(C) Posterior changes in persistent avalanche activity are dominantly represented in alpha band activity while frontal changes are dominantly represented in theta activity.}
\label{fig:frequency_dependence}
\end{figure*}

\end{document}